\documentclass[aps,superscriptaddress,amsmath,amssymb,floatfix,twocolumn,showpacs,amsfonts,longbibliography,pra]{revtex4-1}
\usepackage{times}
\usepackage[varg]{txfonts}
\usepackage{textcomp}
\usepackage{graphicx}
\usepackage{subfigure}
\usepackage{tabu}
\usepackage{color}
\usepackage[colorlinks=true,citecolor=blue,urlcolor=blue,linkcolor=blue,hyperindex]{hyperref}
\usepackage{braket}
\usepackage{float}
\usepackage{overpic}
\usepackage{gensymb}
\usepackage{mathrsfs}
\usepackage{tikz}
\usepackage{bm}
\usepackage{multirow}
\usepackage{amsfonts}
\usepackage{amsmath}
\usepackage{mathrsfs}
\usepackage{amssymb}
\usepackage{paralist}
\usepackage{indentfirst}
\usepackage{ulem}
\usepackage{soul}
\usepackage{cancel}
\usepackage{CJK}

\begin{document}
\title{Interplay of Unidirectional Quantum Strings in Kagome Rydberg Atom Array}
\author{Wei Xu}
\affiliation{Department of Physics, and Chongqing Key Laboratory for Strongly Coupled Physics, Chongqing University, Chongqing, 401331, China}

\author{Xue-Feng Zhang}
\thanks{Contact author:  zhangxf@cqu.edu.cn}
\affiliation{Department of Physics, and Chongqing Key Laboratory for Strongly Coupled Physics, Chongqing University, Chongqing, 401331, China}
\affiliation{Center of Quantum Materials and Devices, Chongqing University, Chongqing 401331, China}

\begin{abstract}
Leveraging the rapid development of quantum simulators, the intriguing phenomena of quantum string are observed across various quantum simulation platforms. 
However, the complex interplay between the quantum strings cannot be well analyzed due to the limited system size in real quantum simulators. 
Here, with the help of a newly developed quantum Monte Carlo method, we can simulate a larger-scale Kagome Rydberg atom array, providing an ideal playground for studying quantum strings. 
By introducing a novel edge pinning method, the ends of a quantum string can be attached to edges so that the flexible manipulation of the quantum string becomes possible.
Due to the geometric constraint, the quantum strings are unidirectional, which strongly complicates their interplay. To quantitatively describe the quantum string, we built a one-dimensional effective model. With both analytic and numerical methods, rich physics can be found, including ``geometric breaking", heart-like superposition state of quantum strings, and the attractive inter-string interactions. 
This work can benefit the comprehension of quantum strings and may also shed light on the simulation of high-energy physics.

\end{abstract}
\maketitle

\textbf{{\textit{Introduction}}}--- Field lines can depict the fundamental interactions between particles, e.g., electric field lines between a positron and an electron, and the gluon string between quarks. 
On the other hand, due to the interplay between geometry frustration and the strong interaction between spins, a line-type topological structure, also known as a 'string', can emerge in quantum frustrated magnetism~\cite{string_review, TIM_string01, TIM_string02, TIM_string03, XXZ_string, kagome_zhang3,string_xiong,dirac_string,spinice,ice_string,qice_string}. 
Interactions between the strings can result in various exotic topological phases of matter \cite{XXZ_string,TIM_string03,TIM_string02,lgt_changle,kagome_zhang2,kagome_zhang3}. 
Unlike the classical Dirac string in classical spin ice \cite{spinice,dirac_string}, here the strings are quantum, meaning the ground state is a superposition of all possible classical string configurations. 

\begin{figure}[t]
	\centering
	\includegraphics[width=0.48\textwidth]{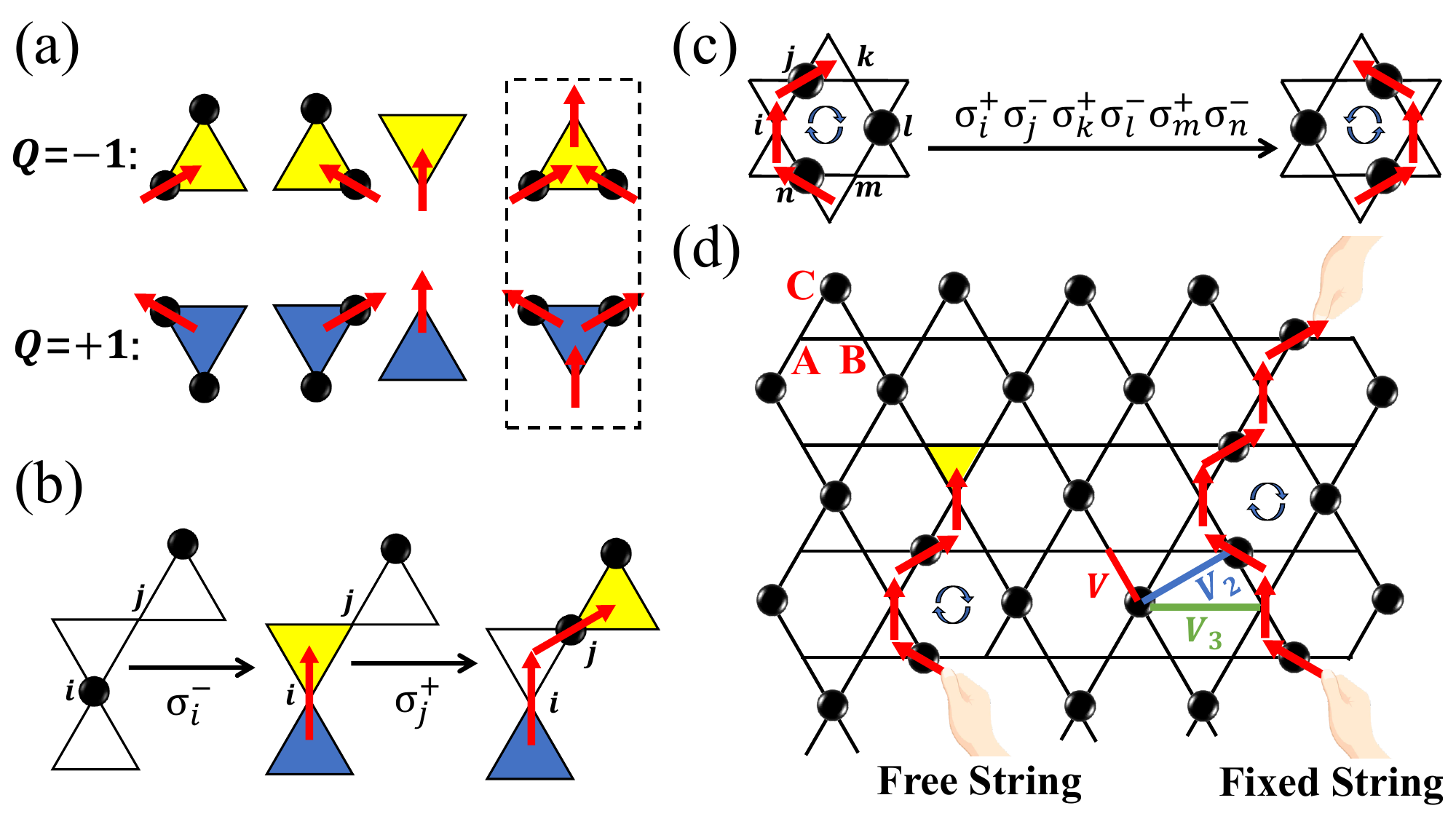}
	\caption{\label{fig01} The schematic diagram of (a) different gauge charges (yellow: negative; blue: positive) with special ones highlighted with dashed rectangle and ``electric field" marked with red arrows, (b) the effect of $\sigma^x$ including creating and moving gauge charges, (c) the ring exchange process highlighted by the recycling symbol arrow, and (d) the ``free" and ``fixed" quantum string (connected red arrows) pinned by the edge defect (hand) in the background of the stripe phase. The repulsive interactions $V_{ij}$ are highlighted by different colored solid lines.}
\end{figure}

As a typical frustrated system, the Kagome lattice is composed of a corner-shared structure \cite{NIS_2012,kagome_Senthil,kagome_wessel,YCH14,YCH15_1,YCH15_2,kagome_zhang1,kagome_zhang2,kagome_zhang3,Rydberg_glass1,Rydberg_glass2}. 
The repulsive (or antiferromagnetic Ising) interaction can impose a geometric constraint analogous to the ice rule \cite{Pauling1935}, so that the disordered ground state preserves macroscopic degeneracy. 
Thus, its low-energy physics is described by the $U(1)$ lattice gauge theory \cite{YCH15_1,YCH15_2,kagome_zhang2,kagome_zhang3}, and the elementary excitations are gauge charges which violate the strong local constraint (Fig.\ref{fig01}(a)). 
If quantum fluctuations or long-range interactions are introduced, the degeneracy will be lifted, so the system becomes ordered \cite{kagome_vbs1,kagome_vbs2,kagome_Senthil,kagome_wessel}, such as the stripe phase with one sublattice fully occupied. As demonstrated in Fig.\ref{fig01}(b), the local quantum fluctuation, like spin flipping, can create/annihilate/move gauge charges. 
Then, the field line (connected red arrows) between the gauge charges is just the string.
Meanwhile, the six-order perturbative interaction (see Fig.\ref{fig01}(c)) can provide a ring exchange which can deform the shape of the string configuration and lower the ground-state energy.
However, it is hard to observe a quantum string due to the confinement of the gauge charges.
Thanks to the rapid development of the Rydberg atom array \cite{Rydberg_Many-body,Rydberg_Single,Rydberg_chain,Rydberg_longr,Rydberg_nature1,Rydberg_nature2}, the spin 1/2 degree of freedom can be encoded with the ground and Rydberg states of the cold atom. Meanwhile, the Rydberg repulsive interaction plays the role of Ising interaction, and the Kagome geometry can be directly constructed \cite{Rydberg_kagome}. Then, the flexibility of the geometry and in-situ manipulation make it possible to prepare and observe the dynamics of the string excitation \cite{Rydberg_kagome,ion_string_breaking,qbit_string_breaking}. 
However, as mentioned in the outlook of Ref.\cite{Rydberg_kagome}, analyzing the interplay between multiple quantum strings becomes an even more complex task, which requires a larger system accommodating more strings.

In this manuscript, we numerically simulate a large-scale Kagome Rydberg atom array with open boundary conditions mimicking the real experiment. 
As shown in Fig. \ref{fig01}, the quantum string is unidirectional and with both ends attached with opposite gauge charges.
With the help of the edge defect, the gauge charge can be pinned so that the one end of the quantum string can be tweezed at the edge (hand in Fig. \ref{fig01}(c)).
Therefore, according to the number of ends fixed, the quantum string can be classified into two types: one-end-fixed and both-end-fixed, which we term as ``free" and ``fixed" quantum strings, respectively (a both-end-free string is not stable due to the confinement of the gauge charges). 
Then, after analyzing two quantum strings with the same or different types by numerical simulations, we find rich physics reflects the matter-antimatter property of the gauge charges, but also the exotic interplay among the strings and gauge charges.

\textbf{\textit{Model}}--- The Rydberg atom array provides an ideal quantum simulator with fine tunability, high flexibility, high precision, and excellent scalability \cite{Rydberg_Many-body,Rydberg_Single,Rydberg_chain,Rydberg_longr,Rydberg_nature1,Rydberg_nature2}. 
Individual Rydberg atoms (e.g. $^{87}$Rb) can be trapped using optical tweezers and arranged into defect-free arrays. 
The array is programmable with the help of the spatial light modulator and acousto-optical deflectors, so the geometry of the lattice can be adjusted with very high flexibility. 
The ground state $\ket{g}$ (e.g. $|5$S$_{1/2}\rangle$) can be coupled to the Rydberg state $\ket{r}$ (e.g. $|70$S$_{1/2}\rangle$) via a two-photon process (e.g. lasers at 420 nm and 1,013 nm) with site-independent tunable Rabi frequency $\Omega_i$ and detuning $\Delta_i$. 
This optical addressability enables exquisite quantum control over the quantum many-body system. Atoms in the Rydberg state feel the long-range repulsive interaction with a Van der Waals (VdW) type form $V_{ij}=V/R_{ij}^6$, where $V=C_6/a_0^6$ if $R_{ij}$ chooses lattice spacing $a_0$ as the units. Then, the strength of $V$ can be adjusted by changing the lattice spacing or the principal quantum number. 
These strong interactions are a hallmark of Rydberg physics, reaching macroscopic scales and facilitating complex collective phenomena. 
Then, the Hamiltonian can be expressed as follows:
\begin{equation}
H = \sum_{i<j} V_{ij} n_in_j-\sum_i \Delta_i n_i-\sum_i  \frac{\Omega_i}{2}\sigma^x_i, \label{eqn01}
\end{equation}
where $\sigma^x_i=\ket{g}_i\bra{r}_i+\ket{r}_i\bra{g}_i$ denotes the exciting process and $n_i=\ket{r}_i\bra{r}_i$ is the density operator of atom in Rydberg state. 
The model above can be mapped into the extended quantum transverse Ising model by implementing the following transformation $n_i\leftrightarrow(\sigma^{z}_i+1)/2$, and the hardcore boson degree of freedom of Rydberg atom $|g\rangle$ and $|r\rangle$ are mapped to the spin half $|\downarrow\rangle$ and $\uparrow\rangle$, respectively. Then, the VdW potential term plays the role of Ising interaction, and the Rabi frequency and detuning correspond to the transverse and longitudinal magnetic field. When taking $\Delta=V$, the leading contribution of the diagonal term can be rewritten as $\frac{V}{2}\sum_k(\sum_{i\in \triangle_k}n_i-1)^2$, which indicates only one atom in each triangle can be excited to the Rydberg state. As shown in Fig.\ref{fig01}(d), to eliminate the next-nearest-neighbor repulsive interaction $V_2$, the degeneracy is lifted so that the ground state changes into the stripe phase.

{\textit{Quantum String}}---It is better to first briefly introduce the lattice gauge theory before discussing the quantum string. Similar to the spin-1/2 XXZ model \cite{kagome_zhang1,YCH14,YCH15_1,YCH15_2,kagome_zhang2,kagome_zhang3}, after subtracting the background field (stripe phase), the ``electric field" is defined as: $E_{ll'} = n_{i}$ ($i\in A, B$) and $n_{i}-1$ ($i\in C$), at the bisector of the corner-shared triangle, where $l$ labels the triangles that compose the honeycomb dual-lattice.
The positive direction is set as pointing from the down triangle to the up triangle. Then, we can immediately find that the triangle rule is indeed the ``Gauss's law", and the violated configurations have different charges calculated by summing over all field lines around one triangle (as listed in Fig. \ref{fig01}(a)). 
Therefore, the stripe phase can be taken as ``vacuum" which has zero electric field.
Meanwhile, gap energies of different gauge charges are approximately equal at $\Delta=V$ and can be tuned by changing the detuning strength \cite{kagome_zhang3}. 
As shown in Fig. \ref{fig01}(d), the non-zero electric field between the gauge charges forms a line-structure with connected arrows which we term as ``string", same as various systems \cite{string_frankpollmann,string_pan}. 
Notice that the gauge charges highlighted in Fig.\ref{fig01}(a) are very special and can attach three ``electric" lines.

As demonstrated in Fig. \ref{fig01}(b), besides the creation and annihilation of the gauge charge pair (like ``vacuum fluctuations"), the Rabi term $\sigma^x_i$ can drive gauge charges hopping between the nearest neighbor sites in the dual-lattice.  
Meanwhile, a sixth-order perturbation can introduce a ring exchange process (${\tikz[scale=0.8]{\fill (0,-0.1) circle (0.5mm);\fill (0,0.2464) circle (0.5mm);\fill (0.3,0.0732) circle (0.5mm);\draw (0,-0.1) -- (-0.1,0.0732) -- (0,0.2464) -- (0.2,0.2464) -- (0.3,0.0732) -- (0.2,-0.1) -- (0,-0.1);}}\leftrightarrow\tikz[scale=0.8]{\fill (-0.1,0.0732) circle (0.5mm);\fill (0.2,0.2464) circle (0.5mm);\fill (0.2,-0.1) circle (0.5mm);\draw (0,-0.1) -- (-0.1,0.0732) -- (0,0.2464) -- (0.2,0.2464) -- (0.3,0.0732) -- (0.2,-0.1) -- (0,-0.1);} $) in each hexagon without breaking the triangle rule \cite{kagome_vbs1,kagome_vbs2}, so that the string can be deformed and all the classical string configurations can be superposed. Then, we can find that the hopping and ring exchange interactions can contribute to the kinetic energy of the gauge charge and the quantum fluctuation of the string, respectively.

The stripe background is strongly disrupted along the quantum string, which means additional energy cost needs to be considered and is proportional to the length, so it is named as ``tension" energy (details in the Appendix.
For a small Rabi frequency, the fluctuation energy cannot overcome the tension energy, so the gauge charges are confined. In contrast, deconfinement is expected at a larger Rabi frequency. 
In the following, we set $\Omega=0.36$ to make sure the gauge charges are confined in the bulk, and the free string with finite length is also allowed.

\textbf{{\textit{Results}}}---The QMC method we utilized is the newly developed high-performance large-scale algorithm \cite{sse_loop}, and the open boundary conditions in both directions are taken to mimic the real experiment (details of the algorithm in the Appendix. The shape of the edge is designed as shown in Fig.~\ref{fig01}(d) so that the outermost sites are fully occupied due to the edge effect. The nearest-neighbor repulsive interaction $V$ and the detuning $\Delta$ are set to one, and the VdW long-range interactions are truncated to the third nearest-neighbor site \cite{Rydberg_glass2,Rydberg_kagome}. We define the length of the first row in sublattice C as $L_x$, and the number of rows in sublattice C as $L_y$.
The maximum system size achieved reaches $N=561$ atom array sites, ten times larger than the state-of-the-art experiment \cite{Rydberg_kagome}. The temperature is set to $\beta=1/T=100$ to approach the ground state physics. All the results are calculated by taking the average of $128$ independent Markov chains with $10^6$ samples in each.

The edge defect can be introduced by changing the local detuning to a large negative value to ensure the atom always stays in state $|g\rangle$.
To highlight the quantum string excitation, we subtract it from the reference stripe state, which means the density in the C sublattice is transformed as $n_i \rightarrow 1 - n_i$. Furthermore, to illustrate the dynamics of the string, we also calculate the density of the resonant hexagon \cite{kagome_zhang1,kagome_zhang2}, which characterizes the strength of the ring exchange process. In all figures, the intensity of the resonant configuration is normalized by the maximum value.

\begin{figure}[t]
	\centering
	\includegraphics[width=0.48\textwidth]{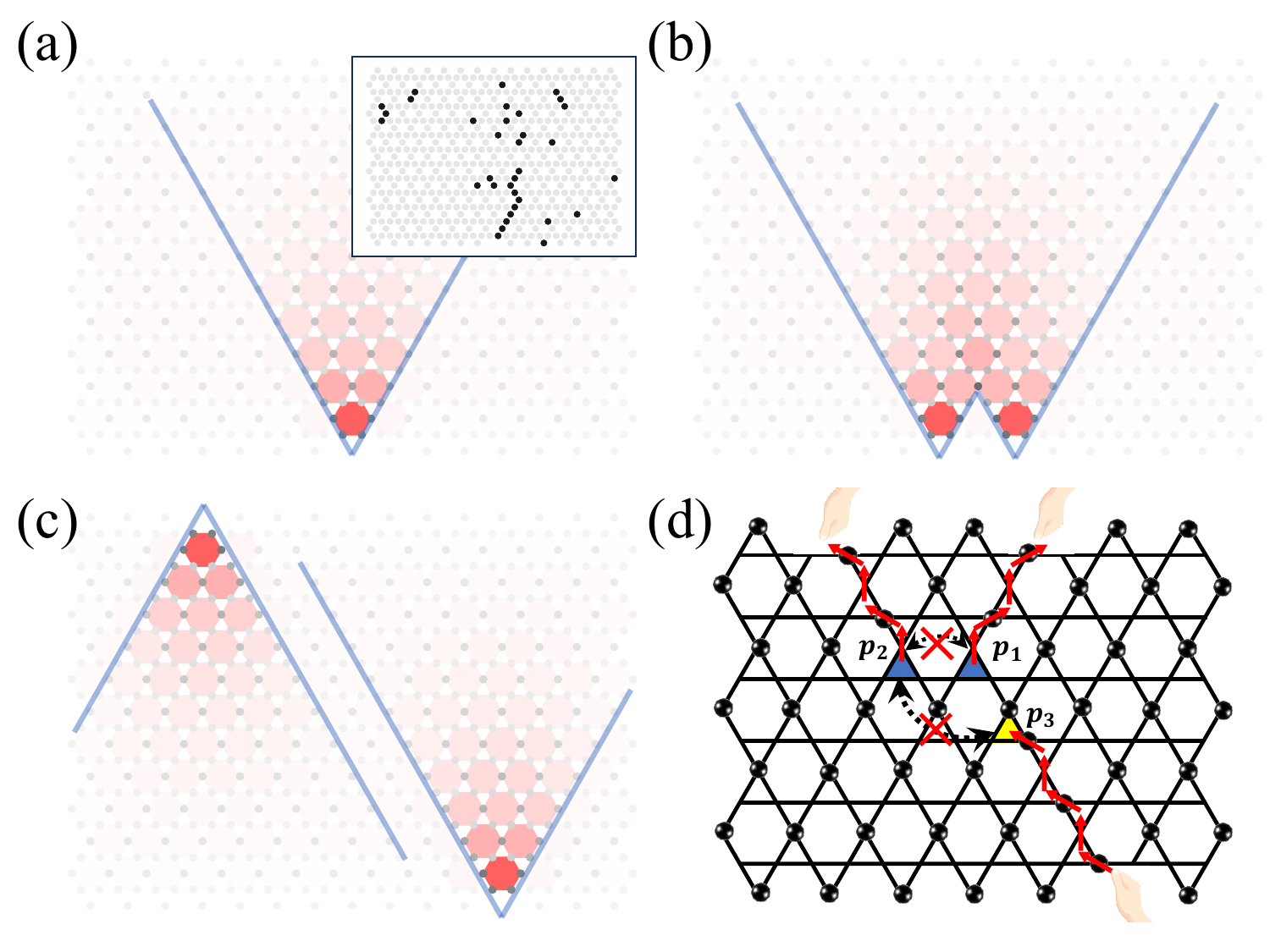}
	\caption{\label{fig02} The density distribution of the atom in the Rydberg state (black dot) and the resonant configuration (red hexagon) for (a) single free quantum string, and double free quantum strings attached with (b) same side and (c) different sides. The blue lines mark the possible region of the string's fluctuation, and the inset of (a) is the snapshot from the QMC simulation after transforming the states in the C sublattice. The number of lattice sites is $561$ with $L_x=15$ and $L_y=13$. (d) Schematic picture of the interplay between free quantum strings, where the red crosses mean the gauge charges can not annihilate with each other.
    }
\end{figure}

As demonstrated in Fig.~\ref{fig02}(a), the densities of both Rydberg excitation and resonant hexagon are higher around the edge defect. The interesting feature is that the sites inside the blue lines labeled region have apparent high intensity, and the strength decays as the site is further away from the edge defect. To figure out the physical reason, we make a snapshot of the transformed Rydberg state distribution. From the inset of Fig.~\ref{fig02}(a), we can immediately observe the string with a finite length attached to the edge. Meanwhile, we find that the string is unidirectional along the $y$-direction, which means it cannot turn around to form a loop-like structure. Such a mobility constraint of the quasi-particle is also observed in the XXZ model \cite{kagome_zhang1} and widely exists in the fracton systems \cite{fracton_review,fracton01,fracton02}. 

\begin{figure}[t]
	\centering
	\includegraphics[width=0.48\textwidth]{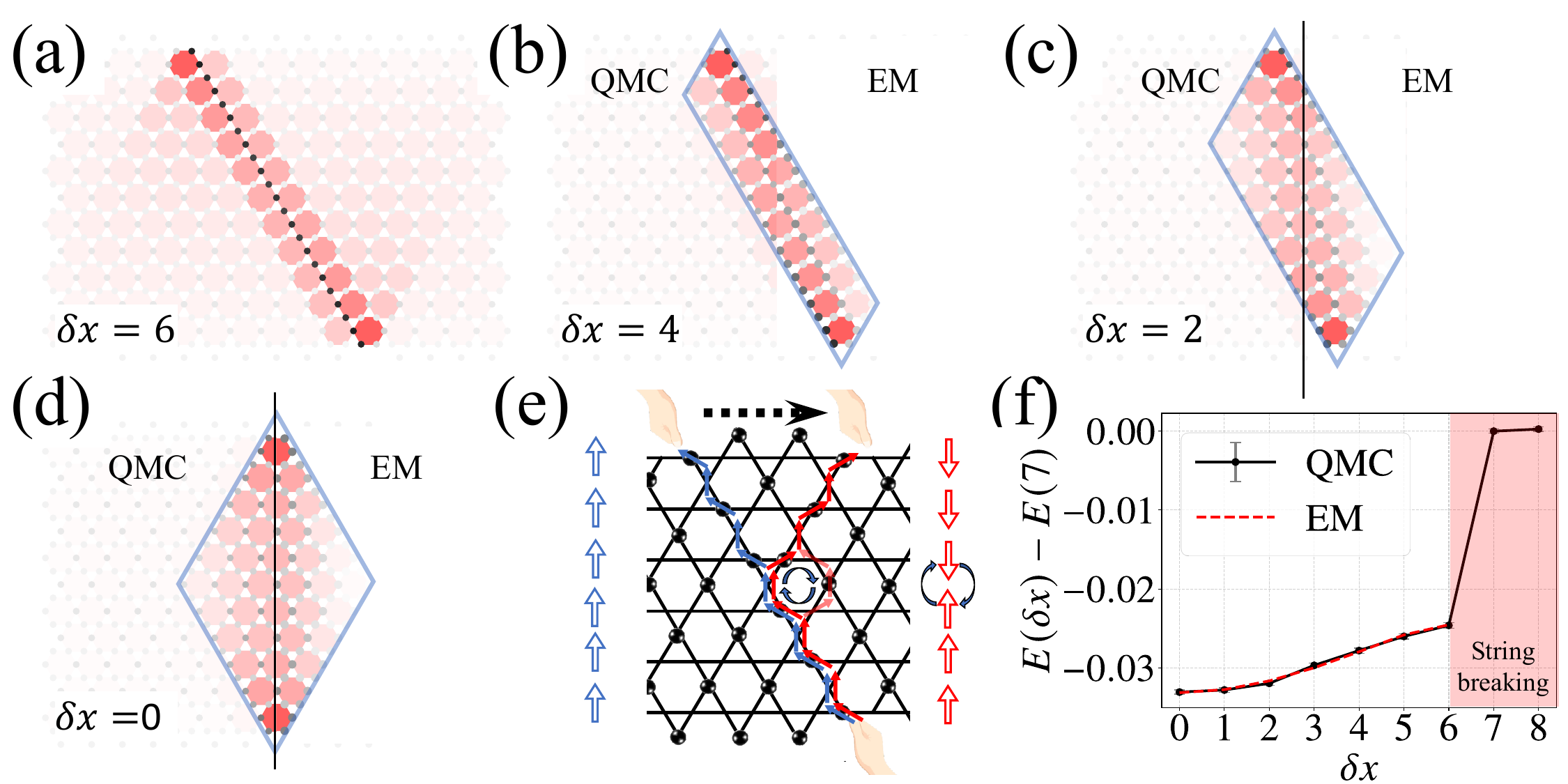}
	\caption{\label{fig03} (a-d) The density distribution of the atom in the Rydberg state (black dot) and the resonant configuration (red hexagon) calculated by QMC (left) and EM (right) for different defect distances with $L_x=15$ and $L_y=13$. (e) The schematic pictures of quantum string excitations presented in (a) (blue) and (d) (red), respectively. The arrows on both sides correspond to the effective spin configuration in the effective theory. (f) The energy difference calculated by QMC and EM, and the energy reference is set to $E(\delta x)$ at $\Delta x=7$, where two free quantum strings are irrelevant. }
\end{figure}

First, we study the interplay between two free quantum strings. If both strings are attached to the same side shown in Fig.\ref{fig02}(b), they can not merge because
the gauge charge on the open ends takes the same charge (see $p_1$ and $p_2$ in Fig.\ref{fig02}(d)). Then, if both strings attach to different sides and stay far away from each other, from the Fig.\ref{fig02}(c), we can see it is also impossible to merge because their fluctuation regions (blue lines labeled) have no overlap even though gauge charges are different (see $p_2$ and $p_3$ in Fig.\ref{fig02}(d)). 
However, if both strings are set to more closer together so that their fluctuation region overlaps, they can merge to a fixed string as demonstrated in Fig.\ref{fig03}.

The distance between edge defects in the $x$-direction can be defined as $\delta x$ in units of $2$. When the defect distance $\delta x$ decreases to $(L_y-1)/2$, two free strings merge into one fixed string, which is straightened and shown in Fig.~\ref{fig03}(a). Along with the decrease of defect distance $\delta x$, the fluctuation region is continuously enlarged as shown in Fig.~\ref{fig03}(b-d). Nevertheless, its shape always maintains a rectangular form. Such characteristics can be understood from Fig.~\ref{fig03}(e). When both ends of the string are pinned, the string length is fixed to $(L_y-1)$ (numbers of arrows $\nwarrow$ and $\nearrow$) because the string is unidirectional and no two segments of the string lie in the same row. Then, the classical configuration with the fewest possible resonant hexagons can be used to define the boundary of the fluctuation region, e.g., the strings in Fig.~\ref{fig03}(e) mark the most left boundary. 

The reverse of the string ``loosing" process is the string ``stretching", and a string breaking can happen for overstretch when $\delta x> (L_y-1)/2$. Different from the ``tensile breaking" in Ref.\cite{Rydberg_kagome}, this ``geometric breaking" is a sudden change in the system which can be reflected in the energies. Thus, we calculate the energies \( E(\delta x) \) with different defect distances \( \delta x \). As shown in Fig.~\ref{fig03}(f), the energy continuously increases while enlarging the defect distances \( \delta x \), which indicates the two defects feel an effective attractive interaction and the quantum string prefers to stay along the $y$-direction. Then, after the defect distance surpasses the limitation $(L_y-1)/2$, the energy undergoes a sudden jump and stays in a plateau ,which indicates that two free quantum strings have no influence on each other after geometric breaking.


As a one-dimensional quantum object, the quantum string can be well described by a one-dimensional quantum model, such as in triangular lattices \cite{XXZ_string,TIM_string03}, Kagome lattices \cite{kagome_zhang1,kagome_zhang3}, and even the high-Tc problem \cite{high-tc}. Illustrated in Fig.\ref{fig03}(e), the configuration of the string can be mapped into an effective spin-1/2 chain by mapping the electric field $\nwarrow$ ($\nearrow$) to effective spin up (down) \cite{kagome_zhang1}. Then, the ring exchange interaction plays the role of the effective spin-exchange or nearest neighbor XY interaction. Meanwhile, the second and third-order long-range repulsive interactions contribute to the ferromagnetic Ising interaction. When the Ising interaction dominates, the quantum string prefers to keep straight so that the stretching will decrease the energy, and the interactions between the defects are repulsive. In contrast, if the ring exchange is dominant, the string would prefer more kinks and stay along the $y$-direction, so the stretching will increase the energy and the two edge defects feel an attractive interaction. Counting the spins in Fig.~\ref{fig03}(e), we can find that the defect distance equals the total magnetization of effective spins. Therefore, zero magnetization corresponds to $\delta x = 0$, and stretching the quantum string is equivalent to tuning the total magnetization. 
Therefore, we can obtain the parameters of the effective model (EM) by fitting the energy with different defect distances. Consequently, as demonstrated in  Fig.~\ref{fig03}(f), the energy calculated by exactly diagonalizing the effective model matches well with the numerical results. Furthermore, as shown in \ref{fig03}(b-d), the distributions of the observables calculated with QMC and EM also match quantitatively well. The corresponding fitted XY interaction is larger than the effective Ising interaction, so we can observe the strong resonances in the middle region. Therefore, similar to the ferromagnetic phase in the spin 1/2 XY chain, the quantum string's dynamics should be more like a free fermion which is also found in the $Z_2$ LGT \cite{borla2025stringbreaking21dmathbbz2}. Details of the effective theory and more comparisons can be found in the Appendix. On the other hand, stretching in the $y$-direction can result in the ``tensile breaking" \cite{Rydberg_kagome}, and more discussions related to distribution and energy are also included in the Appendix.

\begin{figure}[t]
	\centering
	\includegraphics[width=0.48\textwidth]{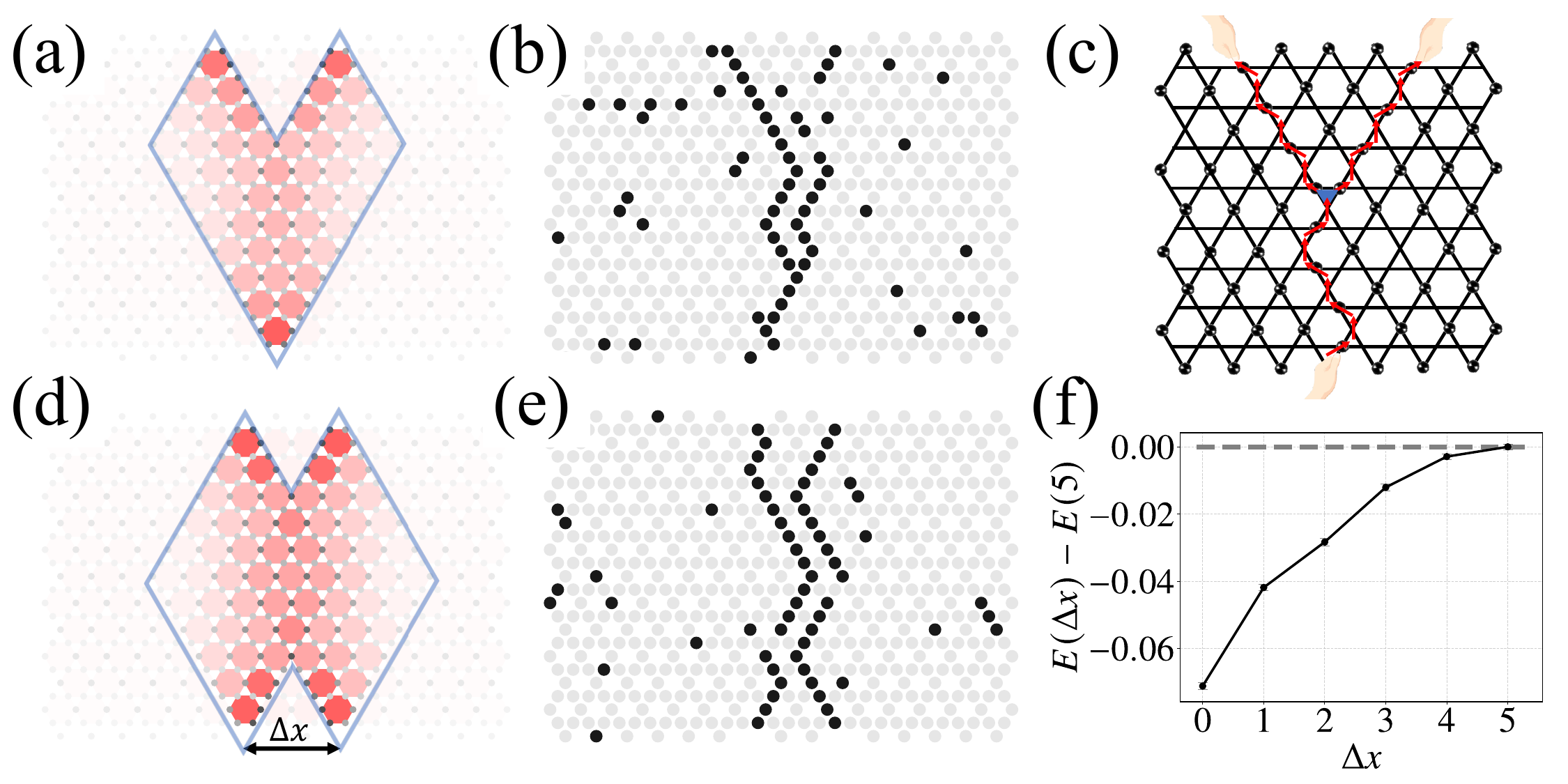}
	\caption{\label{fig04} (a) The density distribution and (b) the snapshot for three defects on different sides. (c) Schematic pictures of the superposition of free and fixed quantum strings. (d) The density distribution and (e) the snapshot for two fixed quantum strings with string distance $\Delta x=2$. (f) The corresponding energy difference of two fixed quantum strings with different distances calculated by QMC, and the dashed black line labels the energy reference $E(\Delta x)$ at $\Delta x=5$, where inter-string interaction is extremely weak.}
\end{figure}

The interplay between the fixed and free strings can be analyzed by setting three defects on different sides. 
Interestingly, as depicted in Fig.\ref{fig04}(a), the density distribution exhibits a heart-like structure with apparently higher density in the middle, which hints that the fixed and free strings are entangled. 
Highlighted in Fig.\ref{fig01}(a), two special gauge charges can attach three electric fields, so they can connect two strings without costing additional energy.
Therefore, as demonstrated in Fig.\ref{fig04}(c), the gauge charge of the free string can attach to the fixed string, and then the ground state can be constructed as a superposition of fixed and free quantum strings. Thus, in the snapshot of the real experiment, the free string is expected to appear on either side of the fixed string, or connect with it as presented in Fig.\ref{fig04}(b).

At last, we would like to check the interaction between two fixed strings. Due to the hard-core constraint of the Rydberg state, the strings can not overlap with each other. Such a no-crossing condition will limit the area of the quantum fluctuation region, so usually the strings feel an effective short-range repulsive interaction \cite{XXZ_string}. However, when introducing the Ising interactions at longer distances, the effective inter-string interaction could be very complex \cite{TIM_string02,TIM_string03,lgt_changle}. Here, both fixed strings are set with defect distance $\delta x=0$, and the distance between the two strings is denoted as $\Delta x$. In the Fig.\ref{fig04}(d), the density distribution of the resonant configuration exhibits high intensity in the middle region, and it indicates the attractive interaction between them. Then, after checking the total energy shown in Fig.\ref{fig04}(f), we can clearly observe that the effective interaction between the fixed quantum strings is attractive. Such exotic phenomena can be understood from the snapshot in Fig.\ref{fig04}(e), where two strings prefer to stay nearby so that the local stripe order with a belt shape can be built in other directions. Thus, noticing the higher density in the middle region from both Fig.\ref{fig02}(b) and Fig.\ref{fig04}(a), we conjecture that such attractive interactions exist between all the quantum strings. 

\textbf{{\textit{Conclusion and Outlook}}}---We analyze the interplay between free and fixed quantum strings by simulating the ground state with the QMC method. If two free quantum strings are attached to different edges, they can merge to a single fixed quantum string when they are close enough. The reverse process can be taken as a novel string-breaking due to geometric reasons, so that the corresponding energy undergoes a sudden change. More strikingly, we find that the free and fixed quantum strings can construct a superposition state reminiscent of the baryon as a bound state composed of more quarks. Finally, we find that the effective interaction between fixed quantum strings is always attractive, which may also be correct for free quantum strings. 

In the future experiment, the complex interactions between quantum strings are expected to bring out rich physics not only in the equilibrium state but also in dynamics. On the other hand, when more quantum strings are considered, the exotic phase with collective behaviors of quantum strings may also emerge. Meanwhile, the quantum strings in higher dimensions are expected to bring out more exotic phenomena, such as the braiding effect. Furthermore, the analysis of the quantum string may shed light on the comprehension of the behavior of the open and closed quantum string in the topological quantum field theory.

\textbf{\textit{Acknowledgment--}}---X.-F. Z. acknowledges funding from the National Science Foundation of China under Grants  No.12274046 and No.12547101, and Xiaomi Foundation / Xiaomi Young Talents Program.

\appendix

\section{Effective model of Quantum string}
It is very complicated to describe the free quantum string, because the length of the string will change when the gauge charge on the free end moves between the dual-lattice sites. However, the fixed quantum string can be well described by a one-dimensional spin model because its length is fixed, and then the interactions in the Rydberg Hamiltonian will produce versatile effective spin interactions.  In this section, we will focus on the effective model of the fixed quantum string, which includes the diagonal tension energy and off-diagonal ring exchange interactions.

\begin{figure}[t]
	\centering
	\includegraphics[width=0.99\linewidth]{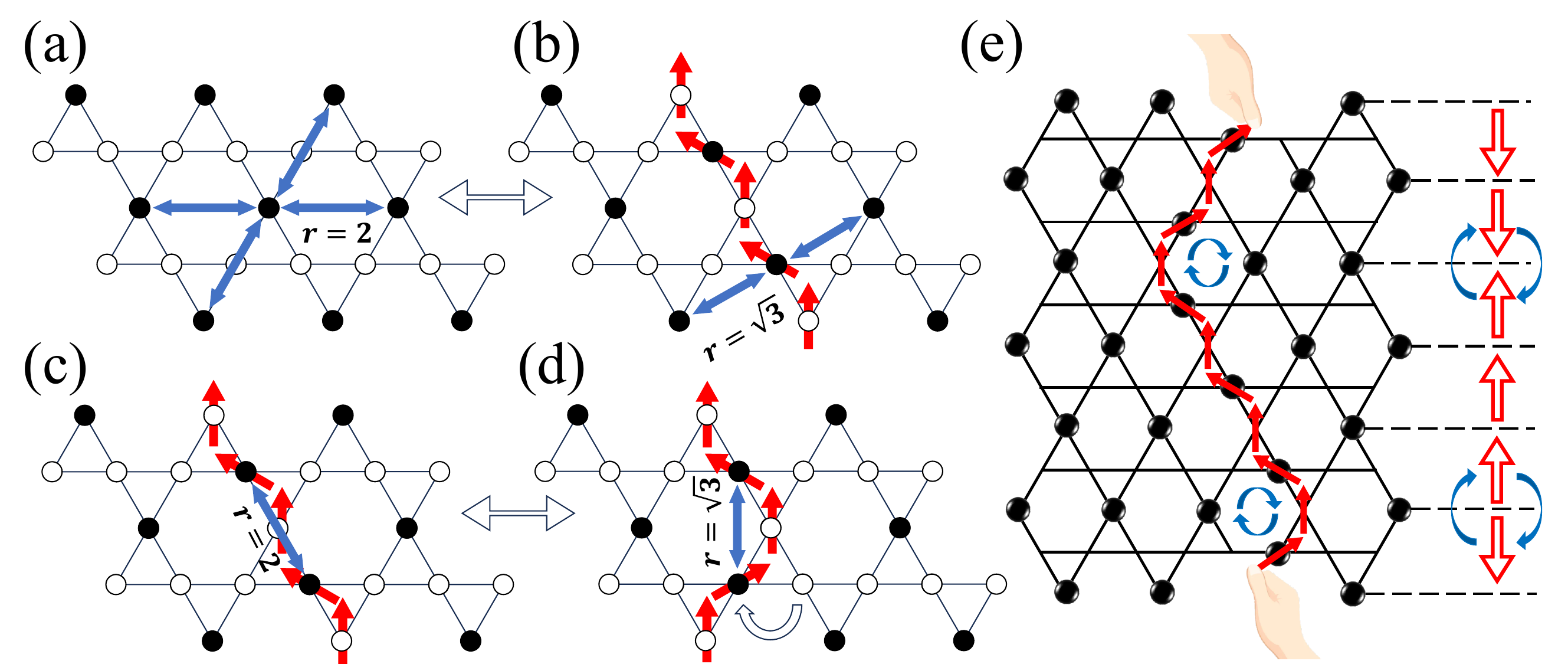}
	\caption{Schematic pictures for analyzing (a-d) the tension energy and (e) the ring exchange interaction of the quantum string. The blue double arrows highlight the Rydberg repulsive interactions considered. White arrows mark the change of positions of excited Rydberg atoms. 
	}\label{figs01}
\end{figure}

The ground state of the Kagome Rydberg atom array is the stripe phase shown in Fig.\ref{figs01}(a), so the insertion of the quantum string will introduce additional energy proportional to the length of the string. Therefore, it is termed as ``tension" energy which results in the confinement of the gauge charges. As demonstrated in Fig.\ref{figs01}, the tension energies are different according to the shape of the string configuration. Comparing with the stripe phase in which all the distances between atoms at the Rydberg state are equal to two, the excited Rydberg atoms along the straight string stay closer to the others in the background as shown in Fig.\ref{figs01}(b). Then, we can immediately calculate that the energy cost per length (or unit cell) is equal to $E_0=V(\frac{2}{3^3}-\frac{4}{2^6})>0$. However, as demonstrated in Fig.\ref{figs01}(c,d), the distance between excited Rydberg atoms along the string becomes shorter if the string is not straight. Thus, the kink on the string gives additional diagonal gap energy $\Delta E=V(\frac{1}{3^3}-\frac{1}{2^6})>0$. Following the mapping process in the main text, the configuration in Fig.\ref{figs01}(c,d) corresponds to $\uparrow\uparrow$ and $\uparrow\downarrow$, respectively. Therefore, the gap energy can be mapped into the ferromagnetic interaction $-JS^z_{i}S^z_{i+1}$ where $J=2\Delta E$.

On the other hand, the six-order perturbative process (${\tikz[scale=0.8]{\fill (0,-0.1) circle (0.5mm);\fill (0,0.2464) circle (0.5mm);\fill (0.3,0.0732) circle (0.5mm);\draw (0,-0.1) -- (-0.1,0.0732) -- (0,0.2464) -- (0.2,0.2464) -- (0.3,0.0732) -- (0.2,-0.1) -- (0,-0.1);}}\leftrightarrow\tikz[scale=0.8]{\fill (-0.1,0.0732) circle (0.5mm);\fill (0.2,0.2464) circle (0.5mm);\fill (0.2,-0.1) circle (0.5mm);\draw (0,-0.1) -- (-0.1,0.0732) -- (0,0.2464) -- (0.2,0.2464) -- (0.3,0.0732) -- (0.2,-0.1) -- (0,-0.1);} $) can deform the string configuration without breaking the local geometric constraint or ice rule. As described in Fig.\ref{figs01}(e), such a high-order term corresponds to the spin exchange interaction or the ferromagnetic XY interaction. In order to maximize this off-diagonal energy, more kinks or antiferromagnetic bonds are required. However, the ferromagnetic Ising interaction or tension energy prefers the ferromagnetic order or fewer kinks. Therefore, these two terms compete with each other. In addition, higher-order perturbative processes can contribute ring exchange in a larger loop, and they correspond to more complex interaction, such as XY interaction in the next-nearest-neighbor site.

Building upon the preceding analysis, the fixed quantum string can be mapped to a XXZ spin 1/2 chain and described with the Hamiltonian:
\begin{equation}
H=-t\sum_i^{L_y-1}(S^+_iS^-_{i+1}+h.c.)-t'\sum_i^{L_y-2}(S^+_iS^-_{i+2}+h.c.)-J\sum_i^{L_y-1}S^z_{i}S^z_{i+1}.
\label{EffectiveModel}
\end{equation}
Here, we set the amplitude of the XY interaction $t$ and $t'$ as the fitting parameters. 

The effective model Eqn.\ref{EffectiveModel} can be directly solved by the exact diagonalization (ED). After obtaining the ground-state wavefunction, each basis vector with the representation of the effective spins can be mapped into the two-dimensional configuration of the Rydberg atom states. Then, the distribution of the atom in the Rydberg state and the resonant configuration can be straightforwardly determined. Furthermore, the thermal effect can also be included by calculating the partition function with full diagonalization. 

\begin{figure}[t]
	\centering
	\includegraphics[width=0.99\linewidth]{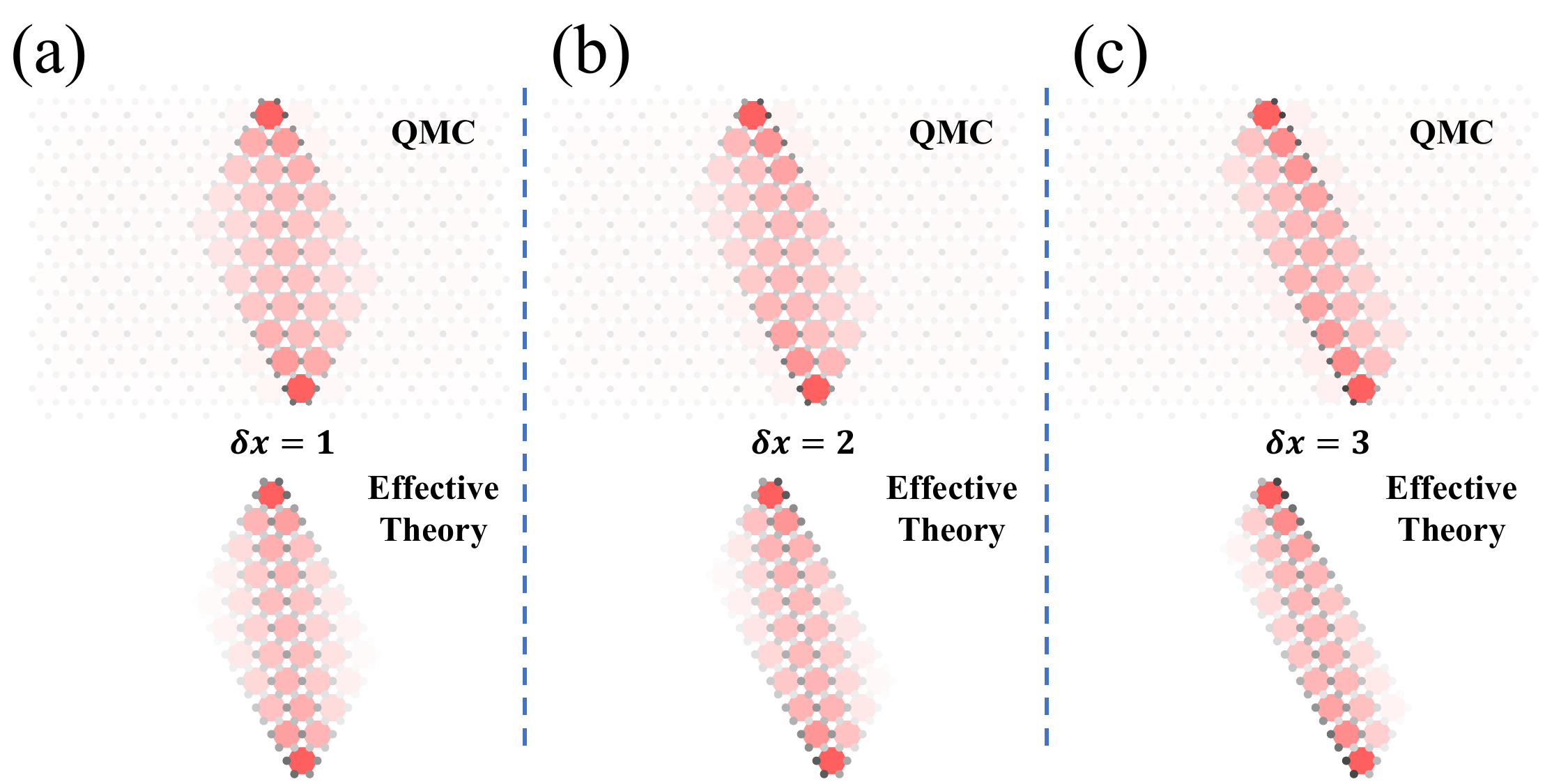}
	\caption{The density distribution of the atom in the Rydberg state (black dot) and the resonant configuration (red hexagon) calculated by QMC and effective theory for (a-c) different defect distances $\delta x$ with $L_x=15$ and $L_y=13$.
	}\label{figs02}
\end{figure}

The distance between two edge defects $\delta x$ is equal to the total magnetization $M=\sum_iS^z_i$ in the effective model. Therefore, by changing the total magnetization, we can simulate the string stretching with the effective model. Then, by fitting the spin exchange coefficient with the energy $E(\delta x)$ at $V=1$ and $\beta=100$, we obtain the best fitting value $t/V=0.022$ and $t'/V=0.0029$. As shown in Fig.\ref{figs02}, the results calculated by QMC and effective theory match very well at different $\delta x$.

\section{Tensile breaking}
\begin{figure}[t]
	\centering
	\includegraphics[width=0.99\linewidth]{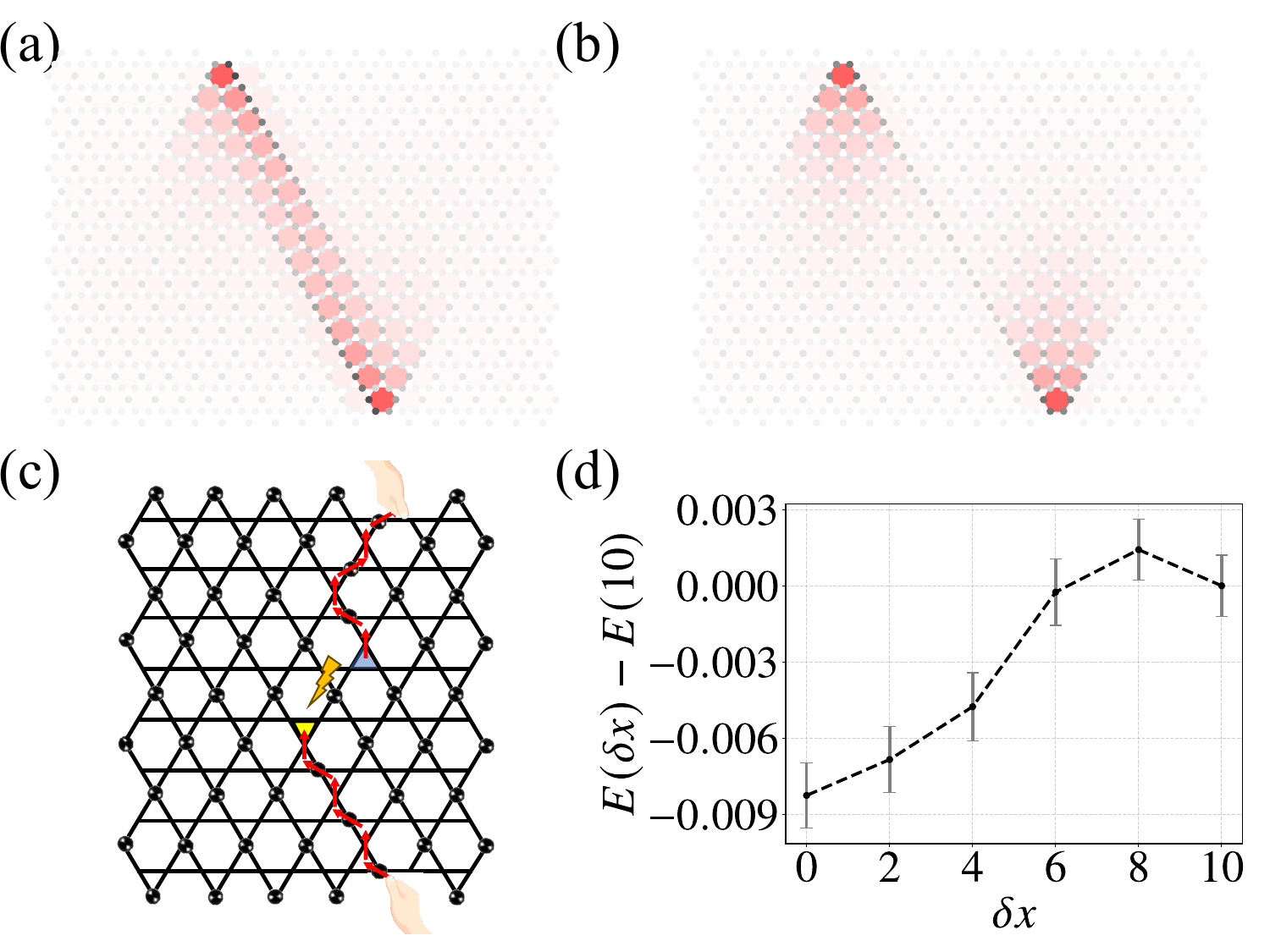}
	\caption{(a,b) The density distribution of the atom in the Rydberg
		state (black dot) and the resonant configuration (red hexagon) for two
		defects in different sides with $L_x=19$ and $L_y=17$ calculated by QMC. (c) Schematic pictures of tensile breaking. (d) The corresponding energy difference.
	}\label{figs03}
\end{figure}

When stretching the string in the $y$ direction by enlarging $L_y$, the quantum string can also be broken because the tension energy ($\sim$O($L_y$)) is larger than the gap energy ($\sim$O($1$)) of gauge charges. Therefore, as shown in Fig.\ref{figs03}(a,b), two free quantum strings can not merge into a single fixed one, reflected by the disappearance of the rectangle fluctuation region. However, we can still observe a blurry string between defects. It means the ground state may be a superposition of broken and unbroken states as depicted in Fig.\ref{figs03}(c). That is why the energy increases at short defect distances and do not show any sudden jump at large $\delta x$ in Fig.\ref{figs03}(d), which is strongly different from the quench dynamics observed in Ref.\cite{qbit_string_breaking, Rydberg_kagome}.

\section{QMC algorithm}
\label{sec:loop}
The Hamiltonian describing the Rydberg atom array is free of the sign problem, so we can utilize the QMC method to simulate it. The method we used is the loop algorithm we developed for the transverse field Ising model in a longitudinal field \cite{sse_loop}, and we would like to review it briefly.

The algorithm is based on the stochastic series expansion (SSE) \cite{SandVik_SSEforQIM, Melko_SSEforRydberg, sse_loop} which expands the partition function as follows:
\begin{equation}
Z=\mathrm{Tr} [e^{-\beta H}]=\sum_s\bra{s}\sum_{n=0}^\infty \frac{\beta^n}{n!}(-H)^n\ket{s}.
\label{eqn01}
\end{equation}
After decomposing Hamiltonian $H=\sum_b({-W}_b)$, the partition function changes into 
\begin{equation}
Z=\sum_s \sum_{n=0}^\infty \frac{\beta^n}{n!} \sum_{b_1,b_2,...,b_n} \bra{s}\prod_{j=1}^{n}W_{b_j}\ket{s},
\label{eqn03}
\end{equation}
so the basic idea of QMC is to sample $\bra{s}\prod_{j=1}^{n}W_{b_j}\ket{s}$ in $d+1$ dimension.

For the Rydberg atom array, the basis vector $\ket{s}$ of the Hilbert space is the Fock state composed of $\ket{r}_i$ and $\ket{g}_i$ at each tweezer site. Meanwhile, the Hamiltonian can be decomposed into:
\begin{equation}
H=\sum_{ij}-B_{ij}-\sum_iO_i-\sum_iC_i,
\label{equ05}
\end{equation}
with
\begin{equation}
B_{ij}=\begin{cases}
-V_{ij} n_in_j+\Delta(n_i+n_j)/z+\delta_{ij} & r_{ij}=1 \\
-V_{ij} n_in_j+\delta_{ij} & r_{ij}>1
\end{cases}
\label{equ06}
\end{equation}
where $\delta_{ij}$ is the energy shift to make sure $B_{ij}$ is positive, and  $z$ is the coordination number, which equals four in the Kagome lattice. Meanwhile, the single-site operators $C_i=\Omega/2$ are the energy shift that can be used to exchange with $O_i=\frac{\Omega}2 \sigma^x_i$ in the off-diagonal update. Usually, for the uniform system, the decomposed operators are the same for all the bonds and sites. However, because we adopt the open boundary conditions in both directions, the decomposed operators have to be nonuniform and position-dependent.

The updating of the SSE algorithm can be divided into two parts: diagonal update and loop update. In the diagonal update, the diagonal operator can be updated by exchanging it with the introduced unit operator, and the algorithm just follows the conventional way. For the loop update, we utilized our newly invented method \cite{sse_loop}, which surpasses the state-of-the-art line algorithm \cite{Melko_SSEforRydberg}.

\begin{figure}[t]
	\centering
	\includegraphics[width=0.99\linewidth]{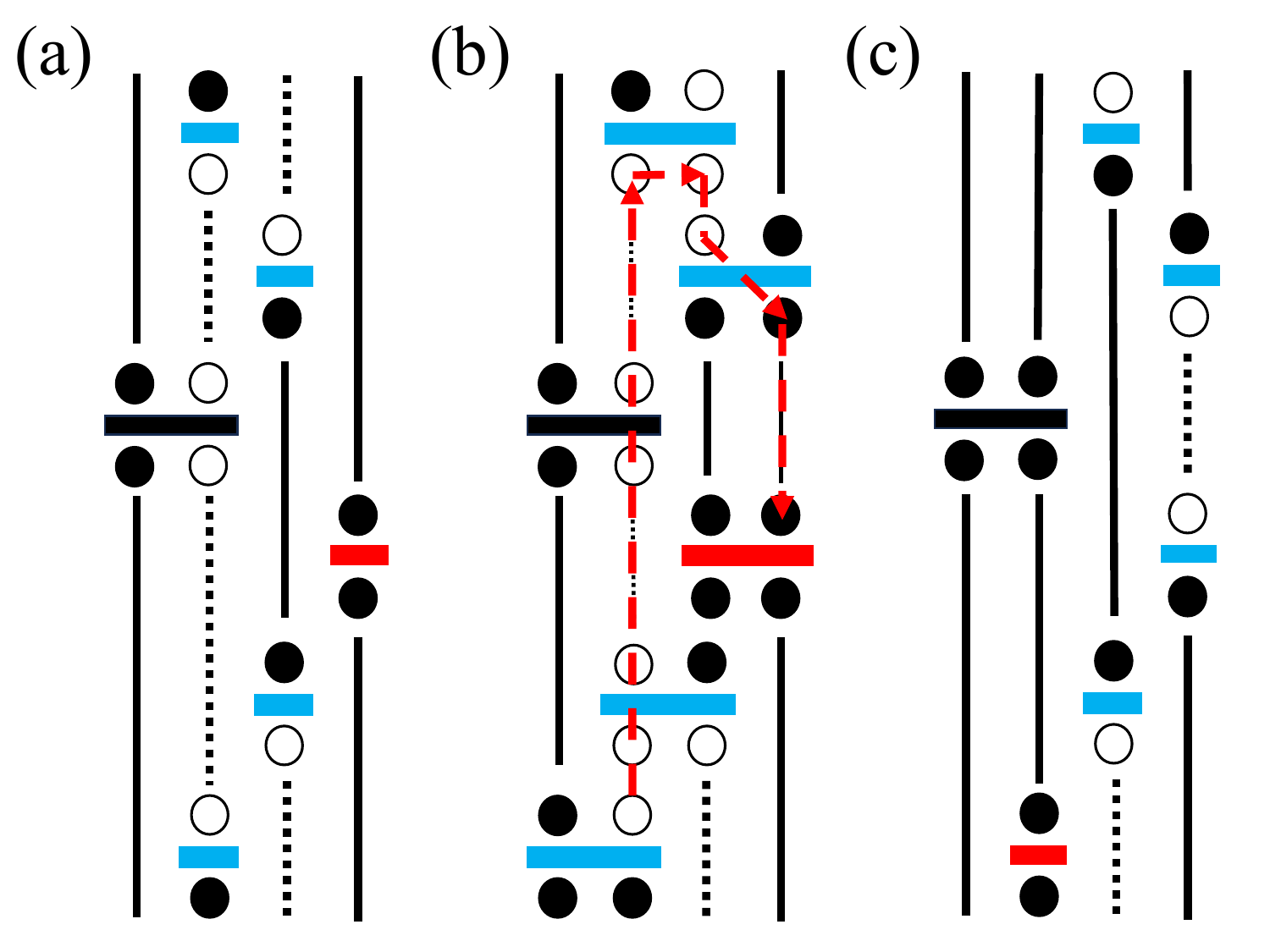}
	\caption{Schematic picture of the off-diagonal update process. Solid lines and black dots stand for Rydberg excited state $\ket{r}$. Meanwhile, the dashed lines and white dots stand for Rydberg ground state $\ket{g}$. Black, blue, and red bars stand for $ B_{ij} $, $O_i$, and $C_i$, respectively. (a) The initial configuration, (b) the loop path after the merge step, and (c) the configuration after flipping spins along the loop path, accompanied by the unmerge process.
	}\label{update_fig}
\end{figure}

The off-diagonal update in our loop algorithm proceeds through three stages: merge, loop, and unmerge \cite{sse_loop}. As shown in Fig.\ref{update_fig} (a) and (b), single-site operators (the constant or off-diagonal operators) first merge with a randomly chosen site to form a merged operator. At the end of the off-diagonal update step, the unmerge process reverts the merged operators back to single-site operators, allowing the off-diagonal operators to move to different sites.

The loop update starts by selecting and flipping a leg. For the constant merged operators, any of the four legs can serve as the starting point. In contrast, for the off-diagonal merged operators, only two legs are eligible for selection; choosing otherwise would result in the emergence of an invalid operator. Then, the spin at the initial leg is flipped, and these two distinct types of merged operators can interchange with each other.

When the loop update path runs on the configuration and flips the legs (red dashed line in Fig.\ref{update_fig}), it will meet three different operators, and corresponding transfer strategies are different as illustrated in Fig.\ref{update_fig}(b). (i) \textbf{Diagonal operator}: only direct passing through and bounce-back are allowed because lack of spin exchange operators, and the acceptability follows the Metropolis way. (ii) \textbf{Off-diagonal merged operator}: the loop update path can randomly exit at one of the other three legs with equal probability. (iii) \textbf{Constant merged operator}:  the loop update path always passes through directly. 

The loop can terminate at a leg of a merged operator. To tune the length of the loop, we introduce a free parameter named the loop-stop probability $P_s$. When $P_s$ equals one, the loop update path will immediately stop upon encountering the first stoppable leg. On the other hand, the loop will never stop at $P_s$=0. In this work, we set $P_s=1/2$. When the loop update path meets an off-diagonal merged operator, it can only stop at the legs with different states. In contrast, any leg of the constant merged operator can be chosen as the ending point, but with an acceptance probability of $P_s/2$. This reduced probability is because the loop update path has only a half chance of encountering the constant operator at the correct position to stop. 

After the termination of these loops, the unmerge step is executed, see Fig.\ref{update_fig}(c). For a merged off-diagonal operator, unmerging involves selecting one side where the spin states are distinct (e.g. $\frac{\circ\bullet}{\bullet\bullet}\longrightarrow\frac{\circ}{\bullet}$). For a merged constant operator, one side is randomly retained. After this step, the off-diagonal update is complete, and measurements proceed as in conventional algorithms \cite{measure}.

\bibliographystyle{apsrev4-1}
\bibliography{referen}
\end{document}